\documentclass{sig-alt-full} 

\newcommand{\ntropy}{{\it N}tropy}

\DeclareFixedFont{\auacc}{OT1}{phv}{m}{n}{12}   
\DeclareFixedFont{\afacc}{OT1}{phv}{m}{n}{10}   

\begin{document}

\conferenceinfo{CLADE'07,} {June 25, 2007, Monterey, California, USA.}

\CopyrightYear{2007}

\crdata{978-1-59593-714-8/07/0006} 

\title{Enabling Rapid Development of Parallel Tree Search Applications}

\numberofauthors{3}

\author{
\alignauthor
Jeffrey P. Gardner\\
\affaddr{Pittsburgh Supercomputing Center}\\
\affaddr{300 S. Craig St.}\\
\affaddr{Pittsburgh, PA 15213}\\
\email{gardnerj@psc.edu}\\
\alignauthor
Andrew Connolly\\
\affaddr{University of Washington}\\
\affaddr{Department of Astronomy}\\ 
\affaddr{Seattle, WA 98195}\\ 
\email{ajc@phyast.pitt.edu}\\
\alignauthor
Cameron McBride\\
\affaddr{University of Pittsburgh}\\
\affaddr{Department of Physics and Astronomy}\\ 
\affaddr{3941 O'Hara St., Pittsburgh, PA 15260}\\ 
\affaddr{cameron@phyast.pitt.edu}\\
}

\maketitle

\begin{abstract}
\vspace{-0.04in} 

Virtual observatories will give astronomers easy access to an
unprecedented amount of data.  Extracting scientific knowledge from
these data will increasingly demand both efficient algorithms as well
as the power of parallel computers.  Nearly all efficient analyses of
large astronomical datasets use trees as their fundamental data
structure.  Writing efficient tree-based techniques, a task that is
time-consuming even on single-processor computers, is exceedingly
cumbersome on massively parallel platforms (MPPs).  Most applications
that run on MPPs are simulation codes, since the expense of developing
them is offset by the fact that they will be used for many years by
many researchers.  In contrast, data analysis codes change far more
rapidly, are often unique to individual researchers, and therefore
accommodate little reuse.  Consequently, the economics of the current
high-performance computing development paradigm for MPPs does not
favor data analysis applications.  We have therefore built a library,
called {\it N}tropy, that provides a flexible, extensible, and
easy-to-use way of developing tree-based data analysis algorithms for
both serial and parallel platforms.  Our experience has shown that not
only does our library save development time, it can also deliver
excellent serial performance and parallel scalability.  Furthermore,
{\it N}tropy makes it easy for an astronomer with little or no
parallel programming experience to quickly scale their application to
a distributed multiprocessor environment.  By minimizing development
time for efficient and scalable data analysis, we enable wide-scale
knowledge discovery on massive datasets.
\end{abstract}

\vspace{-0.02in}
\category{D.1.3}{Programming Techniques}{Concurrent
  Programming}[Parallel programming]
\category{J.2}{Physical Sciences and Engineering}{Astronomy}
\category{D.2.2}{Software Engineering}{Design Tools and
  Techniques}[Software libraries]

\terms{Performance, Algorithms, Design}

\keywords{Parallel development tools, parallel libraries,
  massive astrophysical datasets, data analysis}

\section{Introduction}

This decade will witness the completion of several new
and massive surveys of the Universe. These surveys span many decades
of the electromagnetic spectrum from X-rays (the ROSAT, Chandra, and
XMM satellites) through the optical and ultraviolet (the SDSS, GALEX,
LSST surveys) to the measurements of the cosmic microwave background
in the submillimeter and radio (the WMAP and PLANCK satellites).  They
will also deliver data at a phenomenal rate.  Pan-STARRS and LSST, for
example, will generate several terabytes of data {\em every night}.  At
the same time, simulations of the Universe are becoming larger and
more complex. Even today, a single simulation can use as many as a
billion resolution elements.  The consequence is that data is already
flooding astrophysicists with information. While each of these massive
data sources, both observational and simulated, provide insights into
the formation processes that drive our universe, it is only when they
are combined, by collating data from several different surveys or
matching simulations to observations, that their full scientific
potential will finally be realized.



\subsection{Compute demands of future sky catalogs}

\begin{figure}[t]
\centerline{\psfig{figure=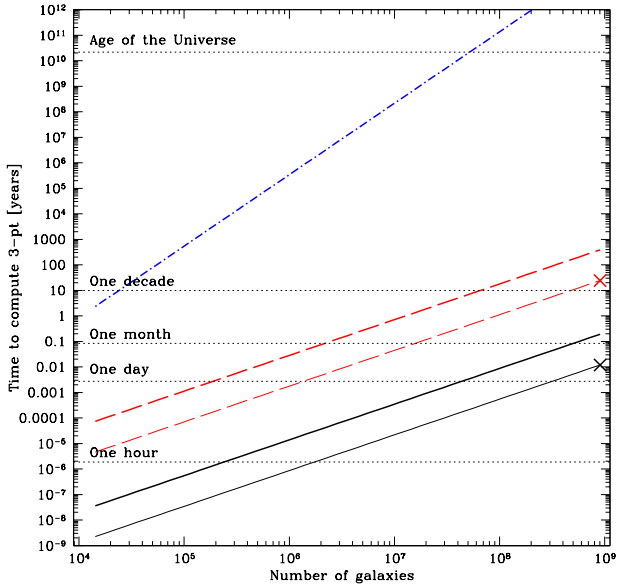,width=0.48\textwidth}}
\vglue-0.25in
\caption{\normalsize Wallclock time required to perform a typical
  3-point correlation function analysis on a dataset of galaxies vs.\
  the number of galaxies in the dataset.  The upper dot-dashed line is
  the naive algorithm that compares every particle combination and scales as
  $O(N^{2.8})$.  The bold long-dashed line is the time required to
  compute using an efficient tree-based algorithm that scales, in the
  best case, as $O(N^{1.4})$ (with the worst case being $~O(N^{1.8}$).
  The bold solid line shows the time required for the same calculation
  on 2048 processors.  The non-bold lines indicate these same
  calculations done in the year 2012, assuming a doubling in CPU
  capability every 18 months.  The stars show the estimated time required to
  process the LSST dataset of 1 billion galaxies in
  2012.\label{fig:execution} } \vglue-.1in
\end{figure}

This unprecedented richness of astrophysical data comes
with an associated challenge. How can an astronomer interact,
visualize and analyze these massive data sets? How can we provide the
user with the ability to pose questions of the data that exploit fully
these new resources?  The next generation of astrophysical surveys
will provide a thousand-fold increase in data rates over the next 3-6
years.  Figure \ref{fig:execution} illustrates the growth in
computational requirements for a common analysis on a galaxy survey
(the exact 3-point correlation function).  Obviously, the use of naive
$O(N^3)$ methods for 3-point analyses are completely out of the
question for modern catalogs like the Sloan Digital Sky Survey (SDSS, \cite{SDSS})
containing 1 million galaxies.  With algorithms based on kd-trees, a
full 3-point analysis is reduced to several weeks of CPU time.
However, by the time LSST comes online in 2012, even accounting for
increases in CPU performance, the same analysis will take over 10 CPU
years.  Clearly, massive parallelism will be required to analyze the
datasets of the future.

There are tools that will allow users with large numbers of small,
independent tasks to quickly and easily distribute their workload to
hundreds or thousands of processors.  However, there will still remain
a class of problems that cannot be trivially partitioned in this
manner.  These are cases where the entire dataset must be accessible
to all computational elements but is so large that it must be
distributed across many nodes.  There are also instances where the
computational elements must communicate with one another during the
calculation. Cluster finding, n-point correlation functions, new
object classification, and density estimation are examples of problems
that will require the astronomer to develop programs for
multiprocessor machines in the near future.  For these problems, the
current tools for analyzing today's data sets will not scale to the
upcoming generation of surveys and simulations.

On the other hand, multiprocessor platforms are becoming increasingly
common.  Perhaps the most notable change to the average astronomer
will be the advent of multi-core processor machines.  Already, CPU
manufacturers are offering 4 cores on a single chip, and this number
will continue to grow over the next few years.  The U.S.\ is
also investing in the cyberinfrastructure required process these data
by steadfastly increasing the processor count of massively parallel
platforms (MPPs) at national resource providers.  Fortunately for us,
the massively parallel supercomputers of tomorrow will be quite
capable of analyzing the sky surveys of tomorrow.  The challenge that
remains is overcoming the hurdles to application development that
prevent their power from being harnessed.

\subsection{Shortening development time for parallel data analysis applications}

Since their introduction in the late 1980s, massively parallel
computers have demonstrated one thing: they can extremely
time-consuming to program.  After climbing the steep learning curve of
parallel programming, the scientist can look forward to spending many
times longer parallelizing their algorithm than it took to write it in
serial.  For this reason, the high-performance computation (HPC)
community is largely dominated by simulations.  Even if it takes 10 or
20 person-years to write a parallel simulation code, the economics
still favor its development since it is typically reused for many
years by many people.  Data analysis, alas, does not work this way.
Every scientist has their own analysis technique.  In fact, it is
largely what makes us each unique as researchers.  For this reason,
astrophysicists do not typically have the time or resources to develop
analysis codes from scratch to run on national compute resources, or
even smaller departmental clusters.  For the full scientific potential
of sky surveys to be realized, we need to create a way to facilitate
the development process of data analysis codes on massively parallel
distributed memory platforms (MPPs).

Procedurally, tree-based algorithms usually employ divide-and-conquer
strategies that are relatively straightforward to parallelize.  The
difficulty for achieving high scalability emerges when the size of the
dataset exceeds the memory capacity of a single computational node,
and A) the tree walks span the domains of many nodes and/or B) nodes
must update data up other nodes as the calculation progresses.  In
these scenarios, which are common in astrophysics, the time required
to communicate between processors bogs the calculation down.  Thus, we
focus on enabling problems in this regime.

Our research has been to design an approach that exploits the fact
that while the number of questions the astronomer may ask of the data
is limitless, the number of data structures typically used in
processing the data is actually quite small.  In fact, most
high-performance algorithms in astrophysics use trees and their
fundamental data structure, and most specifically kd-trees.  This is
because they are typically concerned with analyzing relationships
between point-like data in an $n$-dimensional parameter space.
Therefore, our library, called \ntropy, provides the application
developer with a completely generalizable parallel kd-tree
implementation.  It allows applications to scale to thousands of
processors, but does so in a way that the scientist can use it {\em
without knowledge of parallel computing} thereby reducing development
time by over an order of magnitude for our fiducial
applications.  Furthermore, \ntropy\ is also highly efficient even in
serial and provides a mechanism whereby the scientist can write their
code once, then run it on any platform from a workstation to a
departmental Beowulf cluster to an MPP.  The scale of the computation
is finally set by the scale of the scientific problem rather than the
development time available to the researcher.

\section{Background}

The last 2 decades have seen the development of many parallelization
methodologies.  Our goal in designing \ntropy\ was to take the key
components of a number of existent strategies and combine them into an
intelligent implementation that enabled scientists to write highly
scalable parallel tree-based applications in much less time than it
would have taken them to write the same thing ``from scratch.''
Therefore, the main focus of our research was not necessarily to
create a novel parallel algorithm.  What was
new, however, was the experiment in combining several existing
parallelization strategies under a single umbrella in a manner that
was most useful for a specific target community.  Our motivation for
doing so was based on the success of the N-body cosmology code
``PKDGRAV'' \cite{PKDGRAV}, a highly scalable tree-based gravity
calculator which has been in production for over 10 years and runs
efficiently on a multitude of platforms, from small SMPs to MPPs with
thousands of processors.  PKDGRAV successfully combines several of the
strategies that we will discuss below into a single application.  The
purpose of our research was to see if such a selective deployment
approach could be extended from a specialized astrophysics application
to a more general-purpose parallel tree library that was both highly
scalable and straightforward for scientists to use.

In the following paragraphs, we discuss the pros and cons of several
parallelization techniques.  The literature offers a broad range of
methodologies for efficiently parallelizing data trees, and the ones
that we ultimately chose for \ntropy\ are by no means unique.  The
general difficulty in assessing them is that most have only been
tested on relatively small platforms (8 to 32 processors) whereas we
are interested in scaling to thousands of computational elements.
Thus, we largely restrict our discussion to strategies that are
actively being deployed and benchmarked (with publications) on MPPs
today.  Our goal is to use the lessons they provide to design a
methodology that leverages their strengths while avoiding their
drawbacks.  For the most part, codes on modern MPPs use message
passing libraries, so our first metric evaluating \ntropy\ will be to
approach the high scalability of message passing.  Our second metric
will be to severely reduce the development time of a tree-based
application in comparison to a purely message passing implementations.

\subsection{Agenda-Based Parallelism}
\label{ssec:agendas}

Carriero and Gelernter \cite{Perplexed} divide parallelism into three
conceptual classes: results, agenda, and specialist parallelism.  An
agenda involves a series of transformations that are applied to all
elements of a particular dataset.  Therefore HPC scientific codes
usually fall into the agenda class, since the scientist is operating
in a dataset that is comprised of large numbers of roughly similar
components that interact with one another and undergo a series of
transformations.  Since scientists are our principle target, we will
favor an agenda-like model.

In the topmost layer of an agenda-based application, the programmer
strings together a series of serial commands.  Parallel speedup arises
when certain tasks can be executed simultaneously by all processors.
A simple example is a DO loop.  A loop of $N$ iterations can be
distributed over $N_{PE}$ processors, with each processor calculating
roughly $N/N_{PE}$ iterations, provided that all iterations are
independent from one another.  The advantage is that the programmer
need only write a serial pieces of code, and the compiler takes care
of all of the gory details of message passing and synchronization.

There have been many compilers developed over the years that provide
the programmer with an agenda-based view of the computation.  It would
be impossible to examine most of them, however briefly.
Nonetheless, it is possible to make some generic observations.  One
stumbling block for many was that they were designed for shared memory
architectures, which are relatively rare in the HPC arena these days.
Another general problem is that most have difficulty scaling beyond a
few hundred processors, where they frequently become dominated by the
overhead involved in spawning, synchronizing and releasing
computational threads.  The compiler proceeds along a single thread
until it identifies instructions that can be conducted in parallel,
whereupon it launches the parallel computation, then synchronizes at
the end.  But spawning tasks and synchronizing is
expensive---sometimes these operations can require nearly 1 million
CPU cycles on distributed memory machines (including NUMA systems).
Consequently, the parallel regions must be extremely large, or
``coarse grained,'' in order for the computation to scale.  Designing
an agenda-based compiler is a battle between granularity and
generality.  Granularity is generally determined by the sophistication
of the tasks in the agenda.  

Some agenda-based
parallel compilers attempt to provide higher-level capabilities in
order to increase granularity.  A good example of this strategy is ZPL
\cite{Chaimberlain04}, which provides the programmer with a way to
manipulate an $n$-dimensional shared array in a spatially aware
manner.  One can operate on the array as a whole: e.g.\ shift array
elements along principle axes.  One can also operate upon array
elements conditional to their location in the array: e.g.\ add my
value $x$ to that of my neighbor above me if that neighbor has flag
$f=$TRUE.  By giving the programmer the ability to string together
very high-level array manipulation commands, ZPL enlarges the
granularity of the computation, allowing it to scale.
Provided that your algorithm fits into this
paradigm, ZPL is an excellent solution that may potentially scale to
thousands of distributed processors.  Unfortunately many scientific
applications cannot be expressed using this formalism, and therefore
find ZPL too restrictive.  A common limitation of such high-level
agenda-based approaches is that your problem must map onto the
high-level instructions and structures that the compiler provides.  In
general, efforts thus far have worked well with regular arrays, but
can be exceedingly cumbersome for algorithms that use irregular and/or
adaptive data structures like trees.  Nonetheless, efforts like ZPL
demonstrate that it is possible to scale well using agenda-based parallelism
provided that each high-level instruction in one's agenda maps easily
onto the computation.  We will revisit this observation when we discuss
\ntropy.

General-purpose agenda-based compilers like UPC, Co--array Fortran, 
P++, and HPF \cite{UPC, CoArray, CoArray2, PPlusPlus, HPF} offer
the advantage that they can be used on almost any problem.  They
achieve this generality, however, by giving the programmer a much more
low-level (and thus generic) set of tools.  Programs therefore tend to
become fine-grained quite rapidly, thereby inhibiting scalability.  It
is difficult to find a balance between granularity and generality.
For these reasons, scientific codes using agenda-based parallelizing
compilers are rarely seen running on the MPPs of today.

%

\subsection{Explicit message passing}

Message-passing libraries provide almost limitless flexibility and
generality.  Interestingly, they accomplish this by forcing the
developer to program at an even lower level than any parallel
compiler.  Because the programmer is in control of any interprocessor
communication, he can use his insight into the algorithm to maximize
its granularity and minimize the effect of network latency.  For these
reasons, nearly all applications that run on modern MPPs use
message-passing libraries.  The most common library by far is MPI.

Like most interfaces that offer a high level of control and
generality, the drawback of MPI is that it is time consuming both to
write in and to learn.  Programming is made even more difficult if the
thread domains are decomposed using structures more complex than
regular grids, because it becomes difficult to use MPIs collective
communication facilities.  Furthermore, MPI poses an interesting
paradox: even though MPI enables largely asynchronous execution, the
more synchronous one's approach is, the easier it is to express it in
MPI.  Similarly, as one's algorithm approaches the ideal of few
barriers and lots of asynchronous communication, it rapidly becomes
quite challenging to implement in MPI.  As we will demonstrate later,
we designed \ntropy\ so that it simplifies the process of writing an
asynchronous application with minimal barriers.  In this manner,
\ntropy\ seeks to enable the developer to achieve nearly the same
performance of a ``hand-written'' MPI application but with much less
effort.

\subsection{Remote method invocation}

One intriguing evolution of the explicit interprocessor messaging
paradigm is ARMI, an advanced ``remote method invocation'' library for
C++ \cite{ARMI}.  RMI (or RPC for ``remote procedure call'')
generically refers to a facility for launching procedures or methods
on a remote processing element.  Message passing libraries like MPI
have a data-centric view of communication in that they simply transmit
data from one location to another.  RMI, on the other hand, means that
you pass an executable procedure, usually accompanied by data, between
physical locations.  Note that each approach is essentially
interchangeable: it is possible to package routines such that they can
be passed via MPI (in fact, this is what ARMI does).  Likewise, it is
possible to use RMI to transfer data: if thread $T$ needs to get data
$x$ from processor $P$'s domain, for example, $T$ would invoke a
method on $P$ that would return $x$.  Some advantages of ARMI---which
is written on top of MPI---is that it is conceptually cleaner than
MPI, and it attempts to aggregate multiple remote invocations together
into a single message, thereby reducing communication overhead.
However, programming in ARMI still does not guarantee scalability.
Navigation of adaptive data structures is typically a serial
operation: one looks at a node or level of the structure, then uses
that information to advance to another location, which must then be
acquired.  Therefore, message aggregation does not, in and of itself,
help us in our quest for extreme scalability for tree codes.  However,
we shall see that RMI does offer
a straightforward mechanism for using an agenda-based approach to invoke
{\em one's own functions} on multiple processors (rather than only
those provided by the compiler).



\subsection{Split-phase execution and process\\ virtualization}
\label{ssec:charm}

One compiler that has achieved some important successes in high
scalability is CHARM++, a parallel extension to C++ \cite{CHARM}.  One
of the design goals of CHARM++ is to offer comparable or superior
performance to explicit message passing strategies while also
simplifying the life of the programmer.  Like ARMI, CHARM++ also
treats communication as the process of sending a methods, along with
relevant data, amongst compute elements.  CHARM++ differs from
traditional RMI approaches in that it uses ``split-phase execution'':
once a remote method is invoked, the invoking thread never receives a
return value, nor can it check on the invokee's status.  In order to
accomplish a roundtrip message, for example, object $A$ invokes object
$B$.  When object $B$ completes its RMI, it must then reinvoke object
$A$.  In practice, however, the goal of split-phase execution is not
to facilitate moving the data to where the computation is, but rather
to make it easy to move the computation to where the data is.  In
other words, $B$ simply carries on with the part of the calculation
that needed the remote data and might not report back to $A$ at all.

Split-phase execution complements CHARM++'s second important feature:
process virtualization.  In this paradigm, one typically creates 100
or 1000 times as many virtual compute threads as processors, and the
threads migrate between processors redistributing workload as needed
\cite{LawlorKale01}.  In our example above, when a object $A$ invokes
remote object $B$, it can elect to suspend itself until it is
re-invoked by $B$ allowing another object to execute.
Therefore, a physical processor should always be busy doing productive
work and never have to wait for messages to complete.  This paradigm
has proved successful in the implementation of NAMD\cite{NAMD}, a
molecular dynamics code that scales to thousands of processors.



Experience has shown that CHARM++'s split-phase execution strategy
does not always mask communication overhead, however.  The language is
quite successful when each virtual thread (e.g. a ``patch'' of
molecules in a molecular dynamics simulation) only needs to interact
with a small number of other processors.  In cases like this, the
split-phase execution model of CHARM++ is an advantage to the
programmer.  Many scientific problems, however, demand that each
computational task (e.g.\ a particle in an n-body simulation) access
an large volume of data distributed across a very broad range of
processors.  Attempts to use CHARM++ for tree-based calculations, for
example, became quickly saturated by network overhead because of the
large number of messages that are spawned during the tree walk.  In
the end, reducing the number of messages turned out to be the deciding
factor, not masking them with computation. The way to reduce the
number of messages is to fetch needed off-processor data via a
round-trip communication (as detailed above), then cache it locally
for future requests by other virtual threads.

The problem with using the split-phase execution model for round-trip
communication is that one must design each method in one's application
so that it can be reinvoked at {\em every point} it requires an
element of distributed data.  For most algorithms, this demands
substantial redesign.  Therefore, for certain problems that must use
round-trip data-centric messaging, split-phase execution can make the
program much more difficult to write, not easier.  Since \ntropy\ is
designed for tree walks, it focuses first on reducing network
communication via data-centric messaging, then masking what remains.
Since our goal is to make our library as simple as possible to use, we
do not employ the split-phase RMI model.  Process virtualization, on
the other hand, is very useful concept to keep in mind for load balancing.

\subsection{Globally shared objects in a distributed environment}

Quite a number of compilers and libraries offer the ability to map
distributed data onto a logically global space.  In fact, most
agenda-based compilers in section \ref{ssec:agendas} offer this
capability.  There are other efforts that offer shared address spaces
in a manner that more naturally supports the programmer in their quest
for maximal granularity while retaining flexibility.  Global
Arrays\cite{GlobalArrays}, for example, allows the programmer to
create arrays that are logically global but physically distributed
across processors.  By allowing asynchronous one-sided communications,
Global Arrays gives the programmer the convenience of globally shared
data without an increase in granularity.  Other compilers like
Linda\cite{Perplexed} give the programmer access to a logically shared
``tuplespace'' in which tuples of data can reside in a globally
accessible manner.  The goal of all of these mechanisms is to reduce
the often-substantial burden on the programmer for managing a dataset
that is distributed across many processors.  They
also allow the library of compiler itself to distribute the data
and optimize its communication in a manner that is most appropriate to
the runtime architecture.

We found in the previous section that we wished to enable a
data-centric messaging model for the individual threads of our
tree-walking mechanism.  Therefore, it would be highly beneficial to
put our tree data, which is physically distributed across all
processors, into a logically shared space.  Any mechanism we provide
for interacting with this data can also incorporate performance
enhancements, like data caching and prefetching.  We were not able to
find scaling data for Linda for even 100 computational elements.
Therefore, it was difficult to assess its potential efficacy for
scaling tree-based applications to thousands of processors.  Global
Arrays is powerful if the developer uses arrays directly, although it
can become unwieldy for creating and managing a flexible data structure
like a tree.  Furthermore, we wished to provide in \ntropy\ a
mechanism for accessing shared data that lent itself naturally to a
tree structure: instead of requesting an array index, one would
request a specific tree node.  A node can even be addressed by
requesting the ``parent,'' ``sibling,'' or ``next'' of the current
node.  \ntropy\ also has the capability to globally share arrays of
structures or objects, which is also important for applications that used
particles and trees.  In other words, we sought to offer to developers
who use trees a similar mechanism to what Global Arrays offers to
developers who use arrays.

\section{Methodology}

\begin{figure}[t]
\centerline{\psfig{figure=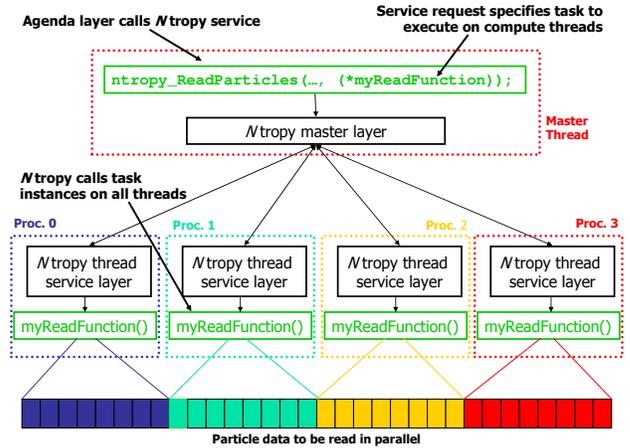,width=0.4\textwidth,angle=270}}
\vglue-0.3in
\caption{\normalsize How to use \ntropy\ to read in a data file in
 parallel.  The user writes the computation ``agenda'' which runs on
 the master thread only and steers the application.  In the agenda,
 they call the \ntropy\ service {\tt ntropy\textunderscore ReadParticles()} in which
 they specify a custom task {\tt myReadFunction()}.  Instances
 of that task are then launched on all threads.}
\label{fig:ntropyReadParticles}
\end{figure}

Algorithms that distribute a tree across many computational nodes have
historically proved to be among the most difficult to parallelize,
because the most effective data structure for organizing the
particles, a tree, is adaptive and irregular.  Moreover, a typical
treewalk often examines many tree cells that are spread across many
processors.  In order to achieve scalability, \ntropy\ employs several
data management techniques like caching of
interprocessor data transfers, intelligent partitioning of the
high-level tree nodes, and dynamic workload management.  All of these
capabilities are time-consuming to write from scratch.
Using the \ntropy\ library, however, the developer gets all of them
for free.  Consequently, we have been able to reduce the time required
to develop scalable parallel data analysis applications by over an
order of magnitude.  Furthermore, many of our \ntropy\ applications
actually perform {\em better} than competing efforts written with much
greater effort from scratch.  Our success proves that it is possible
to build a general-purpose parallel library that is easy to use,
efficient, and scalable.

\subsection{Ntropy structural components}

Fundamentally, \ntropy\ is a library that provides communication and
thread control infrastructure for parallel kd-tree computations.
It incorporates a variety of concepts such as computational agendas,
remote-method invocation (RMI), and message passing.  The strength of
\ntropy\ is that it exploits each of these concepts when necessary and
avoids them when they hinder scalability or usability.

The first piece on any \ntropy\ application is the {\em agenda,} which
serves as a computational steering mechanism and as an RMI launch pad
for invoking parallel subroutines.  An example of an \ntropy\ service
would be to read in data in parallel from an external file and is
illustrated in Figure~\ref{fig:ntropyReadParticles}.  In the agenda,
the user calls {\tt
ntropy\_ReadParticles(..., nParticles, fileName, (*myReadFunction));}.
\ntropy's infrastructure then invokes the method {\tt
myReadFunction()} on all of the compute threads.  \ntropy\ offers
facilities for RMI on both generic and specialized routines.  Our
example uses the specialized interface {\tt ntropy\_ReadParticles()}
that tells \ntropy\ that the method the user is invoking is designed
to read in {\tt nParticles} elements of particle data from file {\tt
fileName}.  This causes \ntropy\ to do a little bit of extra work to
make life more convenient.  Each compute node calculates the indices
of its beginning and end particles and allocates sufficient storage
space.  Then each thread invokes the custom callback function
{\tt myReadFunction(fileName, startPartID, nPartsToRead, ptrToParts)}.
which opens the file {\tt fileName}, forwards to the particle {\tt
startPartID}, reads in {\tt nPartsToRead}, and copies the
data into the location pointed to by {\tt ptrToParts}.  Once all instances of
{\tt myReadFunction()} have returned, the compute threads
automatically signal completion to the master, which then returns from
{\tt ntropy\_ReadParticles()}.  \ntropy's RMI facility makes the
programmer's life much easier by furnishing a simple interface for
coordinating parallel computation.  The beauty of this approach is
that it retains ease of workflow specification inherent in
agenda-based compilers, but also permits customization at the
per-thread level that maximizes the granularity of the computation.

\subsection{Simplifying access to distributed data}

In principle, an RMI interface is general enough to also provide data
transfer abilities.  In practice, however, we have found that
algorithms benefit greatly from a shared-memory view of distributed
data.  In other words, RMI is great for managing the flow of
computation across nodes but, once those computations have been
invoked, it is easier for the algorithm developer if they can be presented
with an interface that makes distributed data behave as closely as
possible to shared data.  Furthermore, it is substantially easier to
achieve high scalability if we treat methods and data differently,
since we can reduce messaging activity through off-processor data
caching (a capability that we discuss later).
For these reasons, \ntropy\ presents a separate, simplified mechanism
for interacting with globally shared data on distributed memory
machines through its ``shared data interface'' or SDI.

SDI supports simple one-sided operations like ``get'' and ``put,'' as
well as some important enhancements.  Any data can be registered into
globally-shared space, including user-defined structures.  The
interface supports both array-like data (e.g.\ an array of floats or
structs) and tree-like data.  Every cell of a shared data tree has a
unique identifier that can be used to retrieve it.  It is also
possible to request the root cell of a tree as well as the parent,
sibling, or children of a specific tree cell.  Allowing \ntropy\ to
manage the tree structure permits a number of performance
optimizations.  The tree is arranged in memory in a depth-first
fashion, meaning that a depth-first tree walk (i.e. one that always
descends the left child first until reaching a leaf node, then
proceeds laterally) would access memory contiguously.  Furthermore,
\ntropy\ mirrors the topmost levels of the tree on all processors.
How many levels are mirrored is selected at runtime and depends on the
available memory.  Thus, because the SDI itself is aware of kd-tree
structure, it not only makes the programmer's life easier, but it also
provides opportunities for optimization.

A further enhancement to normal one-sided communication libraries is
the way that SDI handles cross-processor writing.  Although a
one-sided ``put'' does not in itself require a barrier, such
operations are difficult without some sort of locking mechanism in
order to ensure that several processors do not update concurrently.
The \ntropy\ SDI provides an automatic ``reduction'' mechanism, where
remote data can be updated via any reducible operation in a completely
asynchronous manner.  This capability is discussed in greater detail
in section 3.3.1 which describes the data caching mechanism.


\subsection{Achieving high scalability}

``Underneath the hood'' of \ntropy\ are two capabilities that
substantially increase scalability: interprocessor data caching and
dynamic workload management.  These are features that are
time-consuming for an application developer to implement themselves,
but come ``for free'' when using our library.

\subsubsection{Interprocessor data caching}
When the application developer registers a block of shared data (as
described above) \ntropy\ logically maps that data onto cache lines.
When an {\tt Acquire()} call results in an off processor memory
access, the entire cache line that holds the data of interest is
fetched.  The idea is that if the thread needed one piece of data, it
will likely need the element next to it as well.  Furthermore, future
{\tt Acquire()} calls for that same piece of data will not need to go
off processor because the data will already reside in the cache.  This
mechanism results in fewer than {\em 1 in 100,000} requests for
off-processor data requiring a message to be sent in current \ntropy\
applications.

At the moment, \ntropy\ has two different kinds of caches: read-only
and ``reduction.''  The read-only cache is the simplest: the
application is not allowed to write to shared memory blocks while the
cache is active.  The reduction cache is for data that is updated as
the computation progresses, and is implemented in a non-blocking
manner that requires no locks or other synchronizations, making it
superior to other parallel concurrent-write mechanisms which must
incur penalties to enforce cache coherency.  The only constraint is
that updates to the cache elements must be commutative and associative
(similar to a parallel reduction operation).  When a reduction cache
is registered, the developer provides a reducer function, essentially
the reduction operator, that takes as input the new value and the old
value of the cached element, then returns a single new value.  Nearly
all read-write operations on shared data in scientific applications
can function with within these constraints, and doing so alleviates
all of the inefficiencies introduced by cache-coherency issues.

\subsubsection{Dynamic workload management}
Load balancing becomes increasingly crucial when scaling to thousands
of processors.  Most existing applications for massively parallel
platforms use a predictive load balancing scheme where the
application analyzes and distributes the entire workload before the
computation progresses.  This is a viable scheme for
simulations---which comprise the overwhelming majority of MPP
applications---since simulation volumes tend to have straightforward
geometries, and the load-balancing behavior from the previous timestep
can be used to extrapolate to the next one.  \ntropy\ applications, on
the other hand, frequently have complex geometries (e.g. the Sloan
Digital Sky Survey volume) and, being data
analysis operations, have no concept of a ``time step.''
Consequently, a more advanced and dynamic load balancing scheme was
required.

\ntropy\ provides a facility that will automatically migrating pending
workload between processors during calculation and is based on the
process virtualization concept of CHARM++.  Instead of invoking a
single instance of a method per physical processor, , say ``{\tt
myFunction()},'' the programmer can elect to invoke many more
instances of {\tt myFunction()} than processors.  In the absence of
workload management, all that {\tt myFunction()} needs to know to
identify the work it must do is the thread on which it is executing.
With workload management, however, {\tt myFunction()} needs a
meaningful descriptor so that it can identify the work for which it is
responsible.  Work descriptors can take two forms in \ntropy.  In the
default scheme, \ntropy\ divides the local tree on each processor into
$N_{work}$ pieces (set by a runtime parameter), with each piece being
described by its root node.  In an n-body gravity calculation, for
example, {\tt MyFunction()} might calculate gravity for all particles
in that tree piece's domain.  Each descriptor has an affinity for the
processor that owns the tree piece.  Affinities are expressed in order
that the task is most likely to occur on the processor that has the
most data relevant to it.  In the second scheme, the developer passes
\ntropy\ a list of work descriptors and processor affinities at the
beginning of the calculation.

We have found that a relatively simple workload migration strategy
suffices quite well.  After all the work has been described \ntropy\
queues the work descriptors on the processors for which they have
expressed affinity.  Each processor then calls {\tt myFunction()} for
each descriptor until finds that it will soon run out of work.
At this point it requests more from a central workload manager.  The
thread with the largest remaining workload then donates several
instances of its work to the requester.  An important attribute of our
work-management system is that a processor predicts when it will soon
run out of work and initiates its request before this happens.  The
new work therefore arrives before the current workload is exhausted,
and no time is wasted waiting for new work assignments.  With a
balanced workload, all threads are kept busy throughout the
computation, and the overall time to solution is decreased.  The
obvious advantage of our implementation to the programmer is that it
takes place entirely ``behind the scenes'' within the library itself.
Since the scientist does not have to recast their algorithm in a
manner that supports split-phase execution, \ntropy's workload
managements system is very straightforward and natural to use.

\subsubsection{Performance diagnostics}
Any performance-sensitive application should have diagnostic
facilities for measuring performance and identifying bottlenecks.
Although relatively straightforward in concept, details like timers
and statistics gathering can be time--consuming to write.  \ntropy\
automatically records timing information for each task instance that
executes, as well as for all I/O operations.  Furthermore, the
\ntropy\ API makes custom timers available to the developer, who
simply resets the timers and turns them on and off when appropriate.
All timing measurements are then furnished upon request (to the
desired level of detail) at the agenda level.  \ntropy\ also records
detailed statistics on interprocessor communication and cache efficacy,
making it easy to determine how much an application is being affected
by communication latency.

\section{Results}

\begin{figure}[ht]
\centerline{\psfig{figure=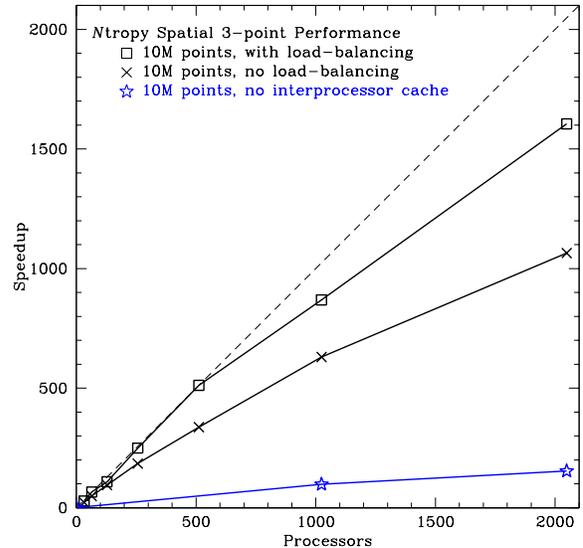,width=0.45\textwidth}}
\vglue-0.25in
\caption{\normalsize The effects on scaling of interprocessor data caching
 and dynamic load balancing.  The open squares show scaling for data
 caching and dynamic load balancing, while crosses demonstrate the
 effects of turning off load balancing.  The open stars illustrate the
 further consequences of disabling the interprocessor data cache.
 This scaling test is for a single spatial 3-point calculation on a
 fixed problem size of 10 million particles randomly distributed in
 the Sloan Digital Sky Survey volume.  It was performed on the PSC
 Cray XT3.}
\label{fig:scaling}
\end{figure}

Two fully functional applications have been written in \ntropy\ so
far: an n-point correlation function calculator and a
``friends-of-friends'' (FOF hereafter) group finder.  Both applications difficult to
parallelize, but for different reasons.  The development time of each
one was reduced by roughly a factor of 10 than if they had been
written ``from scratch'' in MPI with similar performance: from 2 years
to 3 months for n-point and from 8 months to 3 weeks for
FOF.

\ntropy\ was built using the RMI and data transport layers of the
astrophysical n-body simulation code ``PKDGRAV'' \cite{PKDGRAV}.  We
estimate the time required to develop from-scratch MPI n-point and FOF
applications as roughly equal to the time needed to write the same
parallel capabilities into PKDGRAV.  The assumption is that for an MPI
application to achieve the same level of performance as the \ntropy\
n-point and FOF implementations, the necessary excess time would be
roughly the same amount of time it took to write the MPI portions of
PKDGRAV that are used by n-point and FOF.  The development times for
the \ntropy\ implementations reflect how long would be needed for
somebody reasonably proficient with the \ntropy\ library.


\subsection{N-point Correlation Functions}
n-point identifies the number of n-tuples that can be constructed
using particles in the dataset subject to spatial constraints. In
2-point, for example, one is interested in all pairs in a dataset that
can be constructed from particles separated by a distance $d$,
$d_{min} \leq d \leq d_{max}$.  In 3-point, one seeks the number of
triangles that can be made from points in the dataset where the sides
(or angles) of the triangle satisfy certain configurations.
There are two things that make this algorithm difficult to parallelize
efficiently.  Long-range spatial searches can examine lots of
off-processor data, making it extremely latency sensitive.
Furthermore, we are interested in the number of {\em unique} tuples,
meaning that we must search the particles in a particular order,
making the application difficult to load-balance.  \ntropy\ overcomes
these obstacles by substantially reducing interprocessor messaging
with its shared data cache and by automatically balancing the workload
dynamically.  Figure~\ref{fig:scaling} shows the fantastic scaling
that \ntropy\ achieves.  On thousands of processors, it scales 10
times better than the naive case.\footnote{Data points for the naive
case are calculated from cache efficiency measurements of the
cache-enabled runs which track total cache accesses, cache misses, and
time penalty per cache miss.} The complex geometry of observational
datasets such as SDSS prevents static load balancing strategies (which
attempt to predict workload ahead of time) from scaling well.  A
3-point calculation on the SDSS, for example, typically achieves about
50\% ideal scaling on 2048 processors.  Our dynamic load balancing
scheme, on the other hand, automatically migrates work from busy
processors to idle ones as the computation progresses and attains 80\%
scalability for the same calculation.

\subsection{Astrophysical Group Finders}
In a group finder like friends-of-friends, the difficulty is tracking
groups that extend across processor domain boundaries.  First, the
groups of particles are constructed by using the kd-tree for spatial
searchers.  Then, the cross-processor groups are connected using an
iterative graph-based procedure originally designed for shared-memory
machines \cite{SV}.  \ntropy\ enables this algorithm by supporting
user-defined shared irregular data structures like graphs, effectively
mimicking a shared-memory architecture on a distributed machine.  The
shared-memory paradigm is, of course, much easier to program for, and
it offers the scientist a broader choice of algorithms.  For this
reason, development of the group finder was substantially accelerated
by using the \ntropy\ library.

\subsection{Serial performance}
In addition to providing great parallel scalability, we found that the
\ntropy\ version of n-point actually ran 6 to 30 times faster than the
existing widely-used serial implementation ``npt'' \cite{Moore00,
NicholEA06}.  This is because \ntropy\ was written to be maximally
efficient in serial as well.  For example, \ntropy\ arranges the tree
nodes in memory such that a full-depth non-recursive tree walk
(i.e. one that always descends the left child first until reaching a
leaf node, then proceeds laterally) would access memory contiguously.
This makes maximal use of cache and speeds up the tree walk.
Furthermore, each tree node stores pointers to parent, children, and
``next'' nodes.  A ``next'' node is the node to which a tree walk
would proceed if it did not open either child.  Thus, moving from one
node to another requires following only a single pointer.  Thus, by
aggressively minimizing memory accesses, \ntropy\ optimizes tree
navigation and provides a high standard of serial performance.

\section{Conclusion}

We suspect that one reason most previous parallel development
environments and tools have not achieved more widespread acceptance in the
HPC community is that each one provided a single paradigm and forced
every aspect of the application to conform to it.  Our work with
\ntropy\ demonstrates that it is possible to take the effectual
attributes of several parallelization approaches and combine them into
a single facility that offers the developer a range strategies
employable when appropriate.  Specifically, our library provides the
ease-of-use and scalability of agenda-based parallelism while
providing as few constraints as possible on the algorithm by using RMI
concepts to launch user-written subroutines on compute nodes.  These
subroutines are then provided with a shared-memory-like view of the
computation which simplifies programming and enables many
shared-memory algorithms.  Instead of forcing the programmer to adapt
their algorithm to a particular paradigm, \ntropy\ offers several
paradigms each adapted to the needs of the programmer, thereby
providing an intuitive and natural solution to parallel application
development.

\ntropy's selective deployment approach and results also yield useful
insights into the parallelization of data trees.  The single largest
problem faced by distributed tree implementations is communication
overhead.  A tree walk usually traverses a broad range of data and is
largely unpredictable.  If the tree is much larger than local memory,
it is quite difficult to prefetch the data the walk is likely to need.
Strategies like caching are therefore necessary and prove extremely
effective at overcoming communication latency.  A caching scheme can
also efficiently update remote data as well, provided that the updates
can be expressed in terms of a reduction.  \ntropy\ accumulates remote
update directives locally until a cache line is flushed and sent to
the remote node that owns the updated data.  Process virtualization of
divide-and-conquer schemes like tree walks can be highly effective for
load balancing so long as they are implemented on top of a data
caching mechanism.  From an ease-of-programming standpoint, RMI is
useful for providing an agenda-based approach to parallelism that
still gives the programmer the necessary flexibility to implement
their tree walk in a coarse-grained fashion.

\ntropy\ facilitates the use of kd-trees on point-like datasets that
are much larger than the memory of a single computational node.  It
enables the scientist to develop an application that scales to
thousands processors in much less time that it would have taken them
to write a similarly performing application with MPI.  The tree
implementation is also efficient and easy to use even for serial
computations.  \ntropy\ therefore provides a seamless ``upgrade path''
for the researcher allowing them to run their application on any
platform, from their workstation to a massively parallel
supercomputer.  By minimizing development time for efficient and
scalable data analysis, \ntropy\ enables wide-scale knowledge
discovery on massive point-like datasets.

\section*{Acknowledgments}
This work was funded by the NASA Advanced Information Systems
Research Program grant NNG05GA60G and was facilitated through an
allocation of advanced NSF--supported computing resources by the
Pittsburgh Supercomputing Center, NCSA, and the TeraGrid.



\end{document}